# In situ correction of liquid meniscus in cell culture imaging system based on parallel Fourier ptychographic microscopy (96 Eyes)

**AN PAN,**[1,2,3] **ANTONY C. S. CHAN,**[1] **BAOLI YAO,**[2,*] **AND CHANGHUEI YANG**[1,*]

[1]*Department of Electrical Engineering, California Institute of Technology, Pasadena, California 91125, USA*
[2]*State Key Laboratory of Transient Optics and Photonics, Xi'an Institute of Optics and Precision Mechanics, Chinese Academy of Sciences, Xi'an 710119, China*
[3]*University of Chinese Academy of Sciences, Beijing 100049, China*
\* *yaobl@opt.ac.cn* (B. Yao), *chyang@caltech.edu* (C. Yang)

**Abstract:** We collaborated with Amgen and spent five years in designing and fabricating next generation multi-well plate imagers based on Fourier ptychographic microscopy (FPM). A 6-well imager (Emsight) and a low-cost parallel microscopic system (96 Eyes) based on parallel FPM were reported in our previous work. However, the effect of liquid meniscus on the image quality is much stronger than anticipated, introducing obvious wavevector misalignment and additional image aberration. To this end, an adaptive wavevector correction (AWC-FPM) algorithm and a pupil recovery improvement strategy are presented to solve these challenges in situ. In addition, dual-channel fluorescence excitation is added to obtain structural information for microbiologists. Experiments are demonstrated to verify their performances. The accuracy of angular resolution with our algorithm is within 0.003 rad. Our algorithms would make the FPM algorithm more robust and practical and can be extended to other FPM-based applications to overcome similar challenges.

## 1. Introduction

Commercial multi-well plate readers can perform impedance or absorbance measurements on the contents of the wells with 10s per plate and has significant applications in cell culture and pharmaceutical research [1, 2]. But they can only provide gross characterization of the samples by their nature. A next generation imaging-based multi-well plate imager is designed to use a single microscope column to scan either a 96-well or 384-well plate to perform fluorescence, absorption or phase measurements, and is extensively used in large format cell culture experiments. With such a system, individual cells within a culture can be examined for their individual morphology, absorption, thickness, dispersion, integrity, vitality and their connections to neighboring cells. And fluorescence images help visualize chemical compositions and structure and track gene expression in individual cells through specific biomarker methods. However, data throughput rate of such systems is limited by the camera data rate and the mechanical scanning speed of the scanner itself. Commercial 96-well plate cell culture imagers can scan a plate with every 8 mins at 1.2 μm resolution [2], which is around 50 times longer than a non-imaging well plate reader. A high efficiency, low-cost multi-well plate imager with good performance is significant and challenging. Over the last two decades, several novel microscopy methods [3-6] were developed to overcome the space-bandwidth product (SBP) limit of the conventional microscope. We have demonstrated successful high resolution (HR) and large field-of-view (FOV) imaging of the cell cultures in both bright field (BF) and fluorescence mode via on-chip microscopy [3-5]. However, it has an inherent problem that cells have to be grown on top of the imaging sensor, which is completely different from the conventional cell culture workflow and therefore the technology has not found major use.

Fourier ptychographic microscopy (FPM) [6-11] is a recently developed and promising low-cost computational imaging technique with HR, wide FOV and aberration-free quantitative phase recovery by replacing the condenser with a light emitting diode (LED) array and applying angle-varied illuminations. The insight is that the objective can only collect light ranging a certain angle, characterized by numerical aperture (NA). However, parts of the scattering light with higher angle illumination can be also collected due to light matter interaction and the sample's high frequency information can be modulated into the passband of objective lens. Instead of conventionally stitching small HR image tiles into a large FOV, FPM uses a low NA objective to take advantage of its innate large FOV and stitches together low resolution (LR) images in Fourier space to recover HR, sharing the root of phase retrieval [12-16] and aperture synthesis [17]. Herein, no mechanical scanning and refocusing are required to achieve a complex gigapixel image. As such, we collaborated with Amgen and spent five years in designing and fabricating next generation multi-well plate imagers based on FPM. We previously reported a 6-well imagers (Emsight) [18] and a low-cost parallel microscopic system (96 Eyes) that is capable of simultaneous imaging of all wells on a 96-well plate rapidly within cell cultures [2]. The 96 Eyes can achieve high resolution (1.2 μm) BF and aberration-free phase images of 1.1 mm by 0.85 mm per condition at the extended depth-of-field (DOF) of $\pm$ 50 μm within 90 seconds, and single-channel fluorescence images within 30 seconds via 4×/0.23 NA plastic-molded objectives. However, the effect of liquid meniscus on the image quality is much stronger than anticipated, introducing obvious wavevector misalignment and additional image aberration (field curvature and distortion). Different with the larger petri dish in [18, 19], the diameter of each well in our 96 Eyes is 6 mm so that the liquid over the samples cannot be regarded as flat medium. Note that the wavevector misalignment and additional aberration are different between every well respectively due to different conditions. It would be quite challenging to overcome this problem from engineering perspective. We have ever added thick plastic plug to flat the liquid meniscus from engineering perspective, but it failed because it also introduced additional bubbles and aberration in turn due to the inhomogeneous medium and the operation departs from conventional cell culture workflow.

In this paper, we first introduce our 96 Eyes system and the significant meniscus problem in section 2. In section 3, we report an adaptive wavevector correction algorithm (termed AWC-FPM) to tackle the wavevector misalignment and demonstrate its effectiveness, validity and performances. As for one tile, the AWC-FPM algorithm first separates the BF and dark field (DF) images by scanning a 5×5 box and counting the number of BF images. The box with maximum number of BF images will be selected. Next, if the box is not located at center, a global offset of wavevector will be added to all the LED elements due to the meniscus-induced misalignment. And the update order will start from the center of the box and the initial guess will be the up-sampling of the center image. Then a simulated annealing (SA) module [20, 21] will be used to adjust the minor shift of wavevector of each LED individually. Compared with using the SA method directly, our algorithm will be much robust. And the updated update order and initial guess are significant. To verify its correctness, the deviation of each LED is derived. It meets imaginary physical model (cubic function fitting) and matches light tracing results exactly. The accuracy of deviation with our algorithm is within 0.1 mm and the accuracy of angular resolution is within 0.003 rad (0.17°). Further, though bubbles are induced by flatting the liquid meniscus with a plastic plug, parts of tiles can be recovered as the ground truth. And the reconstructions of our algorithm without the plug are coincident with the ground truth and the recovered wavevector distribution is the same as the compensation results of Snell's Law. In section 4, we present a pupil recovery improvement approach to tackle the meniscus-induced additional aberration by utilizing the information of adjacent region, assuming adjacent region shares the same pupil function and demonstrate its performances. The proportion of translation is tested and around 20% is required at least. In addition, we added another channel fluorescence excitation and displayed the performances of the whole 96 Eyes system in Section 5. The throughput of existing commercial products is compared. Finally, we conclude our methods and its potential application space and discuss the future applications of parallel FPM system.

## 2. Working Principle of 96 Eyes and meniscus problems

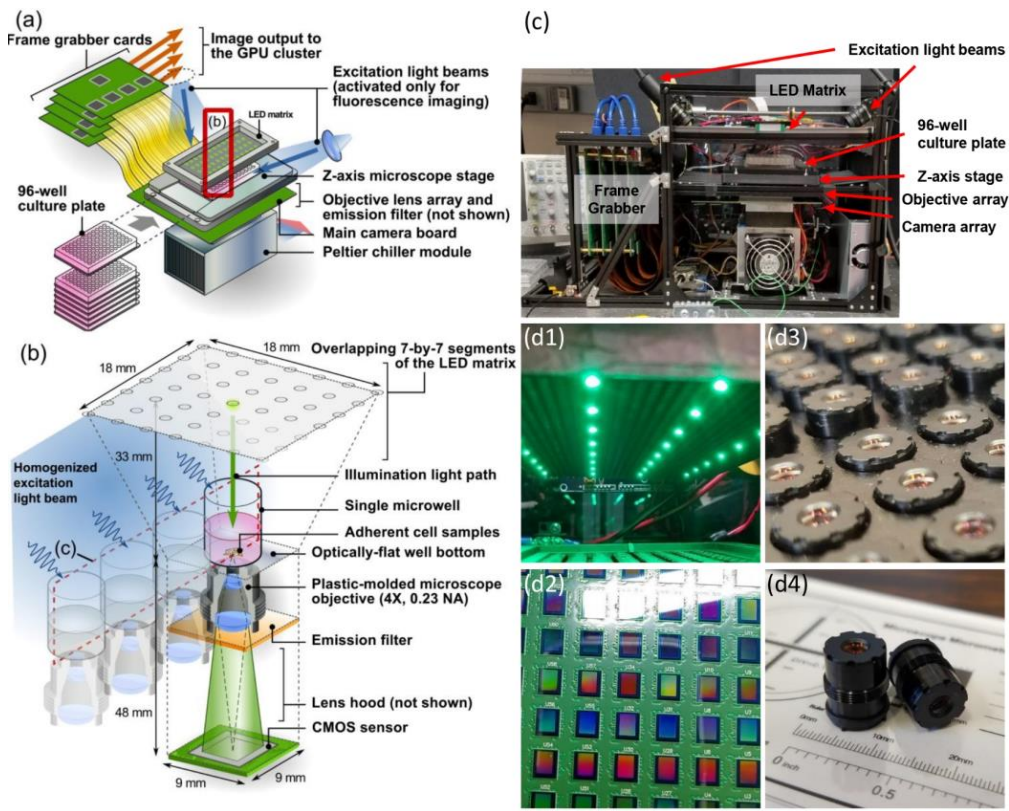

Fig. 1. Schematic and photograph of 96 Eyes. (a) General hardware. Individual plates are loaded from the front. (b) The imaging module consists of 96 repeating units of compact miniaturized microscopes packed in a 9 mm × 9 mm × 81 mm space, where they all share the same light source. (c) The photograph of the 96 Eyes system. (d1-d4) Close-ups of parallel illumination scheme, 96-in-1 image sensor board, objective lens array and objective with external threads of 350 μm pitch, respectively.

The schematic and photograph of 96 Eyes system are shown in Figs. 1(a) and 1(c), respectively. The target 96-well plates are sequentially loaded into the sample stage. A piezo-electric z-axis stage is used to hold the well plate in place and to provide z-axis translation as needed. A programmable green LED array (3mm spacing) is placed at 33mm above the plate with the wavelength of 532 nm. The customized 96-in-1 image sensor board (Fig. 1(d2)) incorporates 96 individual sets of CMOS sensors (1944×2592 pixels, 1.75um pixel pitch) and 4×/0.23 NA plastic-molded objectives (Fig. 1(d3)), each set aligned to the corresponding well of the 96-well plate (Fig. 1(b)). Notably, a finite conjugate optical configuration is chosen for its compactness and mechanical stability, resulting in a fixed object-to-image distance of 48 mm. The external threads of 350 μm pitch (Fig. 1(d4)) allows manual adjustment of lens-to-object distance. The camera sensor board is interfaced with four frame grabber boards, which are controlled by a graphical processing unit (GPU) array in the workstation. A parallel illumination scheme (Fig. 1(d1)) to get rid of all darkfield image recording on the fly is used to captures a batch of 7×7 frames of all the wells on the well plate within 90 seconds. While 25 images are used in practice to get twice better resolution. Different from our previous system [2], two pairs of liquid-guided excitation sources project the homogenized light beam (465 nm and 553 nm) from both side of the culture plate and the emission filter (510 nm and 610 nm) is used to obtain dual-channel fluorescence images. The plate is scanned with a step size of 25 μm (around twice the native DOF) and a range of 100 μm within 30 seconds. In experiments, we used the human bone osteosarcoma epithelial cells (U2OS) as our samples. More details can be found in [2].

Compared with larger petri dish in [18, 19], the diameter of each well in our 96 Eyes is 6 mm so that the liquid meniscus over the samples cannot be regarded as flat medium, which introduces additional distortion, field curvature and wavevector misalignment (Fig. 2). Different height of liquid will result in different degrees of distortion (Figs. 2(a) and 2(b)). Though we tried to keep 200 μL culture solution (10 mm height) for every well in experiments, slight differences can also lead to varying degrees of distortion. Obvious field curvature can be observed from the close-ups of different regions (Fig. 2(d)). Note that locally the field curvature is generally represented as defocus and the distortion is indicated as tilt. This meniscus also brings a challenge of ptychographic illumination, known as the parallax effect, where the position of the light source appears to differ when viewed from different lateral positions on the object plane (Fig. 2(c)). It is common to compensate the wavevector with Snell's Law for the flat medium [18, 19]. But the liquid menisci of different wells have different curvature due to the variation of liquid height and warps of wells. And note that the curvature is also not fixed for different local regions (close-ups in Fig. 2(d)).

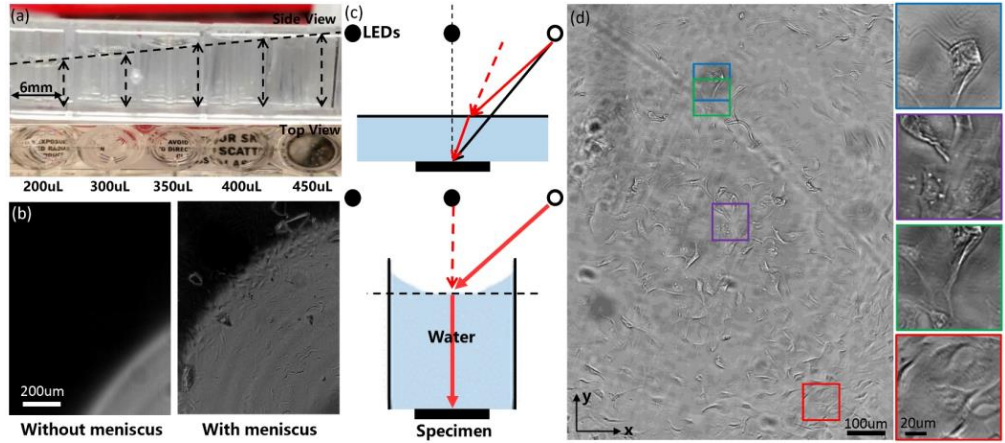

Fig. 2. The distortion, field curvature and wavevector misalignment induced by the meniscus. (a) The distortion with different amount of liquid from the perspective of side view and top view, respectively. (b) The proportion of a typical knife edge image with and without meniscus, respectively. (c) Model of wavevector misalignment with flat and liquid meniscus over the specimen, respectively. (d) Full FOV image of a typical well captured by 4×/0.23 NA and its close-ups. Purple region is in focus, while other regions are in different degrees of defocus. Obvious field curvature can be observed.

## 3. Adaptive wavevector correction (AWC-FPM) algorithm

To tackle the liquid meniscus problem, we reported the AWC-FPM algorithm as follows. First, the BF images are identified after block processing by setting the threshold of 0.1, which is a common strategy to distinguish BF and DF images in FPM [8, 22]. A 5×5 box is scanned to select out the maximum number of BF images (green dash box in Fig. 3). The size of box is set according to the number of BF images, which results from the parameters of system. A proper boundary is set to accelerate the scanning operation, e.g., scanning the 5×5 box within the 7×7 center region in our experiments. The wavevector $k_{m,n}$ of each image in ideal model is set as

$$k_{x,m,n} = k_{m,n}\sin\theta_x = \frac{2\pi}{\lambda}\frac{x_{m,n} - x_0}{\sqrt{(x_{m,n}-x_0)^2 + (y_{m,n}-y_0)^2 + h^2}}$$

$$k_{y,m,n} = k_{m,n}\sin\theta_y = \frac{2\pi}{\lambda}\frac{y_{m,n} - y_0}{\sqrt{(x_{m,n}-x_0)^2 + (y_{m,n}-y_0)^2 + h^2}}$$

(1)

where $(x_0, y_0)$ is the central position of each tile, $x_{m,n}$ and $y_{m,n}$ denote the position of the LED element on the row $m$, column $n$, $\lambda$ is the wavelength and $h$ stands for the height between the LED array and sample. The red dash box in Fig. 3 denotes the BF images in simulated model.

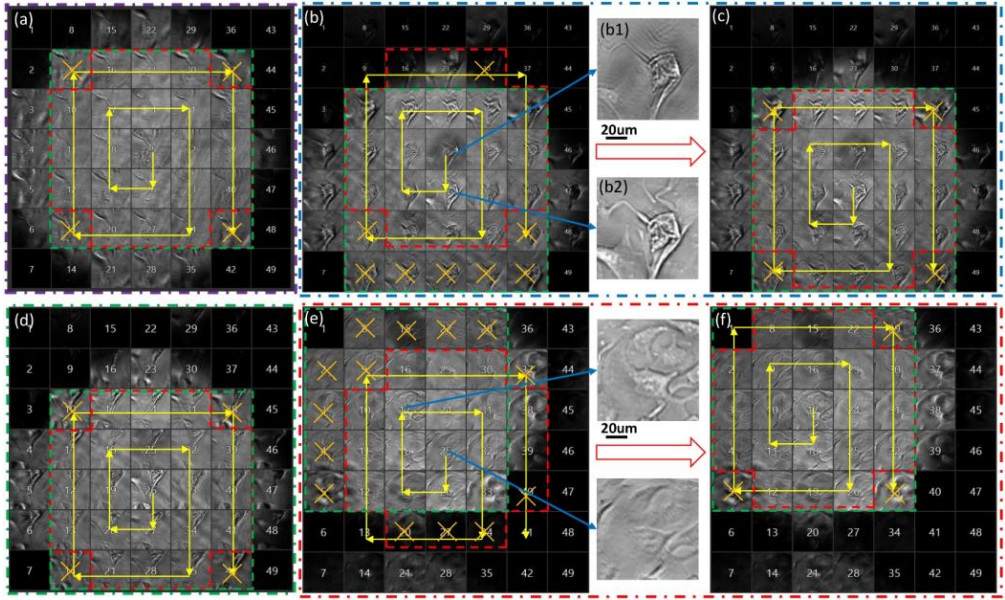

Fig. 3. Illustration of the adaptive wavevector, update order, initial guess and model alignment in AWC-FPM algorithm. (a, b, d, e) Raw dataset of purple, blue, green and red tiles in Fig. 2(d), respectively. (c, d, f) Updated wavevector, initial guess and update order after model alignment. Red dash box: the BF images in ideal model; Green dash box: the real BF images selected by scanning a 5×5 box; Yellow arrays: update order; Orange cross: omitted images due to model misfit (vignetting effect [22]); Serial number: record order.

As for the center tile (purple region in Fig. 2(d)), the red box is coincident to the green box (Fig. 3(a)) except the omitted images (orange cross) due to vignetting effect [22], i.e. the simulated model matches the real case so that the reconstruction is well (Fig. 5). But for other regions, for example, the blue region in Fig. 2(d), the red box does not match the green box (Fig. 3(b)) due to the meniscus-induced wavevector misalignment and results in the wrong discard operation. The reconstruction has many artifacts (Fig. 4(a)). This case indicates a global shift of wavevector. Therefore, a global offset $(\Delta x_g, \Delta y_g)$ will be added adaptively to all the LED elements (red box in Fig. 3(c)) as follows. And the $(\Delta x_g, \Delta y_g)$ can be directly given by calculating the deviation of center images of red and green boxes and is an integral multiple of interval of LEDs.

$$k_{x,m,n} = \frac{2\pi}{\lambda} \frac{x_{m,n} - x_0 + \Delta x_g}{\sqrt{(x_{m,n} - x_0 + \Delta x_g)^2 + (y_{m,n} - y_0 + \Delta y_g)^2 + h^2}}$$

$$k_{y,m,n} = \frac{2\pi}{\lambda} \frac{y_{m,n} - y_0 + \Delta y_g}{\sqrt{(x_{m,n} - x_0 + \Delta x_g)^2 + (y_{m,n} - y_0 + \Delta y_g)^2 + h^2}}$$

(2)

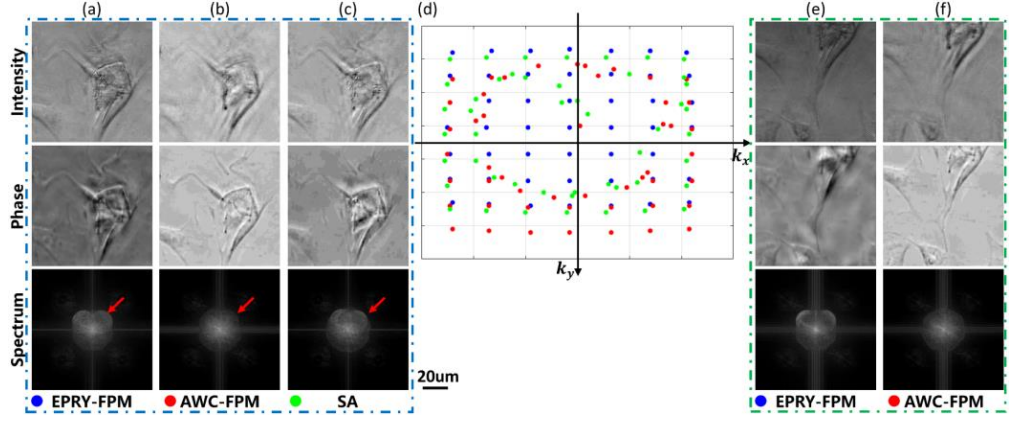

Fig. 4. The reconstructions of blue tile and green tile in Fig. 2(d) with different algorithms, respectively. Results of blue tile with EPRY-FPM algorithm (a), AWC-FPM algorithm (b), SA algorithm (c). (d) Wavevector distribution of different algorithms. Results of green tile with EPRY-FPM algorithm (a) and directly utilizing the wavevector of blue tile for adjacent tile (f).

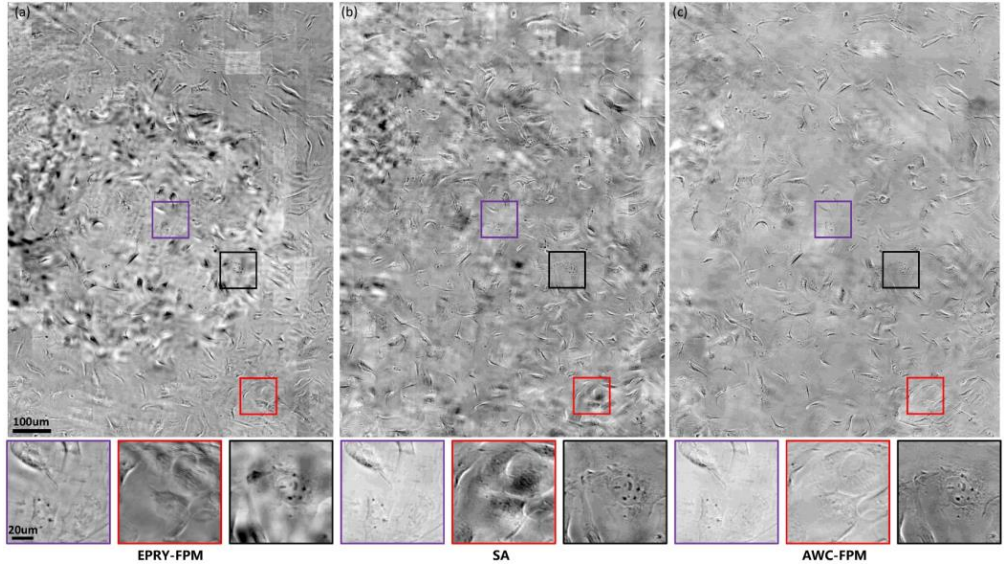

Fig. 5. Phase reconstruction of full FOV with different algorithms. (a) EPRY-FPM algorithm; (b) SA algorithm; (c) AWC-FPM algorithm.

The update order $S_i$, where $i$ is the iteration number, and initial guess will also be updated. Compared the Figs. 3(b1) and 3(b2), Fig. 3(b2) has better contrast and is closer to the result of normal incidence and final reconstruction. And update order will start from the center of the box with a spiral order (yellow arrays in Fig. 3(c)). Identical operations are done to green and red regions in Fig.2(d) as shown in Figs. 3(d) and 3(f), respectively. Note that the global offset is not always in the same direction for all the tiles compared Fig. 3(c) and Fig. (f), because the curvature of liquid meniscus is also not fixed for different tiles, while it may be the same for adjacent tiles (Figs. 3(c) and 3(d)).

Though the global offset has compensated the model misalignment vastly, the wavevector of each LED element is not always an integral shift due to different slope of liquid meniscus and the wavevector is still not accurate. Only adding the global offset cannot adjust the misalignment and the entire FOV reconstruction with global offset only can be seen in Supplementary 1. Next, a simulated annealing (SA) module [20, 21] will be used to adjust the minor shift of wavevector of each LED individually (Fig. 4(b)). A set of further estimates of the wavevector are calculated toward eight directions $r=1,2…,8$. The $r$th estimate of frequency aperture can be given by

$$\varphi_{r,i,m,n}(k_x, k_y) = O_i(k_x - k_{x,m,n} + \Delta k_{r,x,m,n}, k_y - k_{y,m,n} + \Delta k_{r,y,m,n}) P_i(k_x, k_y) \quad (3)$$

where $O(k_x, k_y)$ is the Fourier spectrum of object function, $P(k_x, k_y)$ is the pupil function and each $(\Delta k_{r,x,m,n}, \Delta y_{r,y,m,n})$ is a random frequency-shifting distance between $\pm \Delta_i$ along the $r$th direction. The $\Delta_i$ is the step size of SA module, which is set to 2 (pixel) in our procedure and decreases to 1 after ten iterations. The minor shift of wavevector $(\Delta k_{r,x,m,n}, \Delta k_{r,y,m,n})$ is selected the minimum of the error metric out as follows.

$$E(r) = \sum_{x,y} \left( |\varphi_{r,i,m,n}(x, y)|^2 - I^c_{m,n}(x, y) \right)^2$$

$$R = \arg \min E(r) \quad (4)$$

$$k^u_{x,m,n} = k_{x,m,n} + \Delta k_{R,x,m,n}$$

$$k^u_{y,m,n} = k_{y,m,n} + \Delta k_{R,y,m,n}$$

where $\varphi_{r,i,m,n}(x, y)$ is the simulated wavefront at image plane and is the Fourier transform of $\varphi_{r,i,m,n}(k_x, k_y)$. Compared with using the SA method directly (Fig. 4(c)), our algorithm will be much robust. Because the misalignment is much smaller after global offset and new update order and initial guess match the real case better. Obvious artifacts in Fourier spectrum can be noticed in Figs. 4(a) and 4(c) (red arrow). The recovered wavevector distribution also indicates this conclusion (Fig. 4(d)). The wrong update order of SA algorithm results in wrong distribution of wavevector. Note that only 25 images are used and therefore, only 25 LEDs' wavevectors are updated. Compared with the fixed wavevector in ideal model (blue spot in Fig. 4(d)), the recovered wavevector of AWC-FPM does not shift along one direction globally but expand outside. To verify the effectiveness of wavevector distribution with our algorithm in Fig. 4(d), we use it for adjacent green tile (Fig. 2(d)) and it works well (Figs. 4(e) and 4(f)). The comparison of full FOV phase reconstruction is shown in Fig. 5. An obvious annular artifact can be observed in Fig. 5(a) with original FPM algorithm. The SA algorithm removes parts of the artifacts, but there are still artifacts around the edge of FOV, especially at the top FOV. While no artifacts are in our reconstruction (Fig. 5 (c)) except for some impurities and the background is very clear. The close-ups of each figure also emphasize this conclusion. As shown in the black regions of Fig. 5, subcellular information, e.g. the Golgi apparatus can be observed clearly. Figure 6 shows the detailed procedure of our AWC-FPM algorithm.

To verify the correctness of our algorithm, the deviation $(\Delta x_{m,n}, \Delta y_{m,n})$ of each LED is derived according to the recovered wavevector distribution, which is given by

$$\Delta x_{m,n} = h(\tan \theta'_x - \tan \theta_x)$$

$$\Delta y_{m,n} = h(\tan \theta'_y - \tan \theta_y) \quad (5)$$

where $\theta$ denotes incident angle and $\theta'$ is the updated incident angle and the final wavevector can be calculated as

$$k_{x,m,n} = \frac{2\pi}{\lambda} \frac{x_{m,n} + \Delta x_{m,n}}{\sqrt{(x_{m,n} + \Delta x_{m,n})^2 + (y_{m,n} + \Delta y_{m,n})^2 + h^2}}$$

$$k_{y,m,n} = \frac{2\pi}{\lambda} \frac{y_{m,n} + \Delta y_{m,n}}{\sqrt{(x_{m,n} + \Delta x_{m,n})^2 + (y_{m,n} + \Delta y_{m,n})^2 + h^2}}$$

(6)

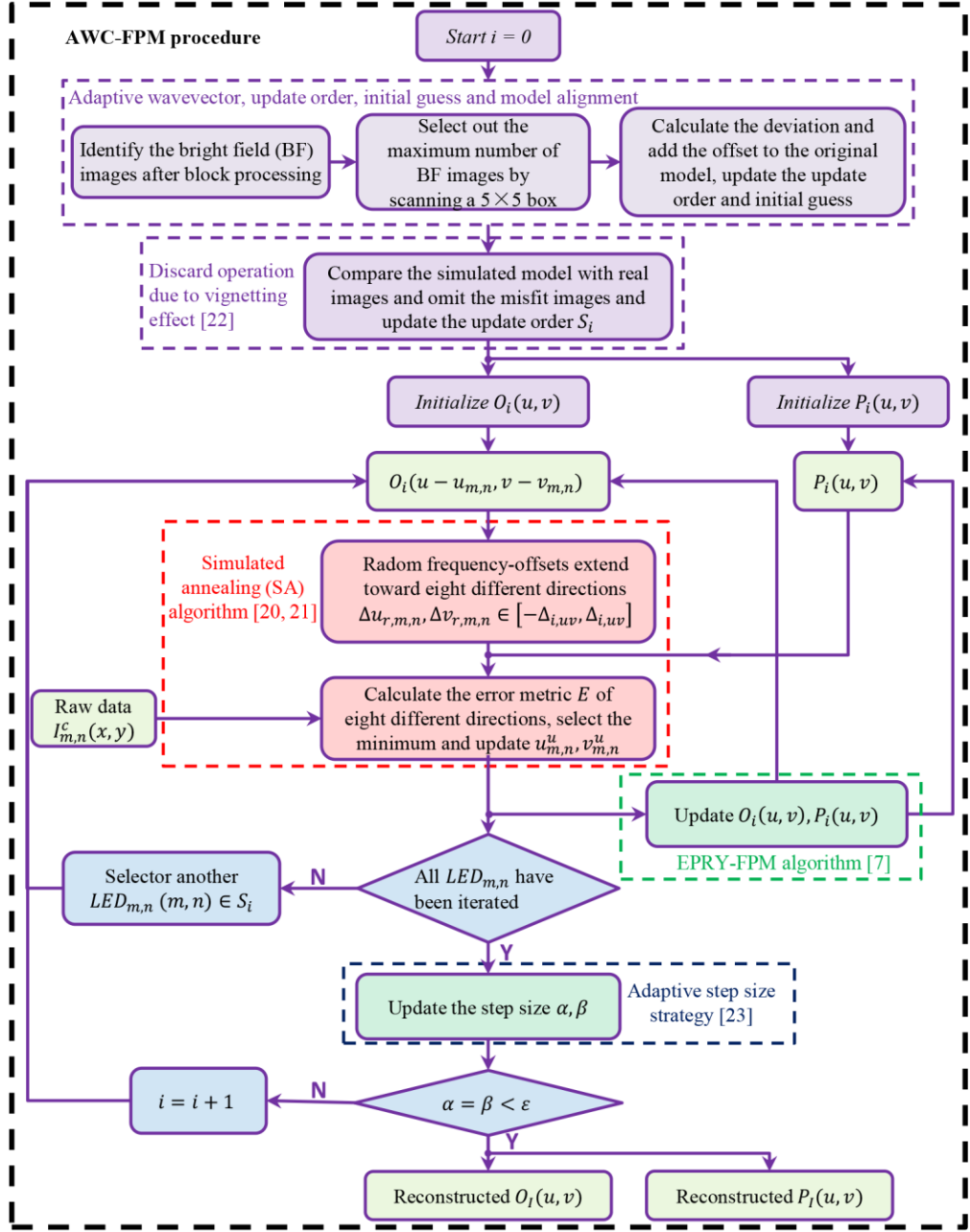

Fig. 6. Flow chart of AWC-FPM algorithm.

The results of center LED are shown in Fig. 7. The x-axis is the center coordinates along x- or y-axis of FOV. The y-axis is the coupled offset we should add. The entire FOV is divided into 23 rows by 17 columns (216×216 pixels per tile) with the overlap ratio of 50% for dense sampling. The dash black line and the solid black line are the theoretical offset, and the practical offset without liquid meniscus, respectively, which is measured by a pure phase Siemens Star. While the blue solid line is the exact light tracing results with liquid meniscus. Our results (red cross) meet imaginary physical model (cubic function fitting, red dash line) and match light tracing results exactly. The accuracy of deviation with our algorithm is within 0.1 mm and the accuracy of angular resolution is within 0.003 rad (0.17°). The purple dash line means when the offset is larger than 3 mm (the interval of LEDs), then the original center will not be the center LED in practice and the updated order and initial guess should be updated as mentioned above. To measure the deviation with light tracing method, a laser diode (532nm) is placed at 33 mm away from the plate and used to scan a well with the same conditions. The step size of translation stage is 1 μm. The bright spot is recorded and a Gaussian fitting is utilized to locate the spot. Then the exact deviation could be obtained by extending the light tracing reversely.

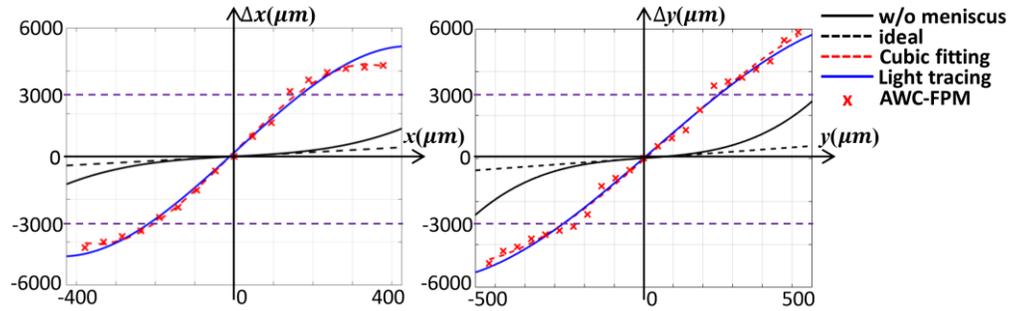

Fig. 7. Derived deviation of center LED with different methods. Dash black line: theoretical offset without liquid meniscus; Solid black line: practical offset without liquid meniscus; Red cross: results of AWC-FPM algorithm with liquid meniscus; Red dash line: cubic function fitting; Blue solid line: exact light tracing results with liquid meniscus.

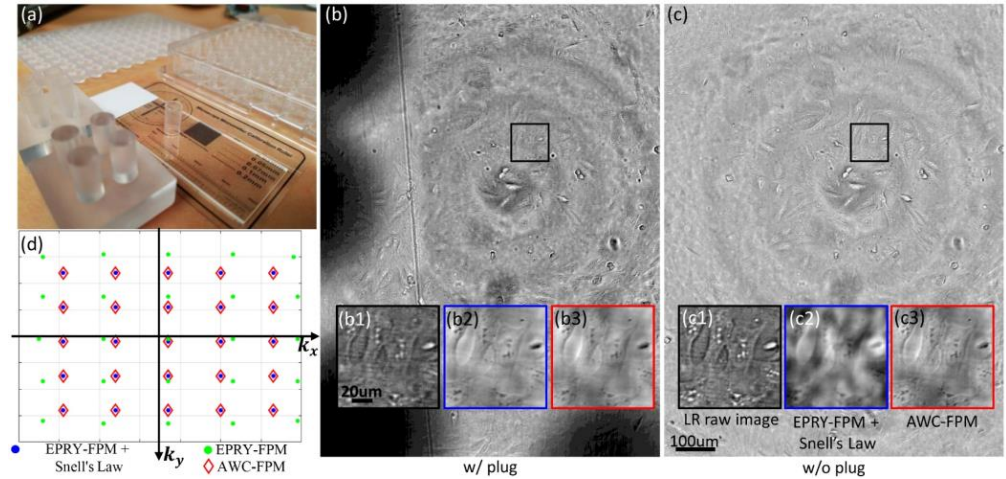

Fig. 8. Reconstructions with or without the plastic plug. (a) Photograph of the plastic plug. (b, c) The full FOV with or without the plug, respectively. (b1, c1) LR Close-ups of the same tile. (b2, c2) Phase reconstructions with original FPM algorithm and compensation of Snell's Law. (b3, c3) Phase reconstructions with AWC-FPM algorithm. (d) The wavevector distribution. Blue spot: compensation with Snell's Law. Red diamond: recovered with AWC-FPM algorithm. Green spot: without compensation.

Further, though bubbles and additional aberration are induced by flatting the liquid meniscus with a plastic plug (Fig. 8(a)), parts of tiles, e.g. Fig. 8(b1), can be recovered with the compensation of Snell's Law as the ground truth (Fig. 8(b2)). The refractive index of the culture solution $n$ is 1.33. The reconstructions of our algorithm (Figs. 8(b3) and 8(c3)) with or without the plug are coincident with the ground truth. The recovered wavevector distribution with our algorithm (red diamond in Fig. 8(d)) is the same as the compensation results of Snell's Law (blue spot in Fig. 8(d)). In fact, we have tried different kinds of plugs (Fig. 8(a)) to flat the liquid meniscus, but the performances are not satisfied and the operation departs from conventional cell culture workflow. Even if we once found a hollowed plug which obtain good performance, but it is hard to fabricate precisely in abundance. Using algorithm to tackle the challenge will be easy to implement and is reliable and low-cost.

## 4. Pupil recovery improvement strategy

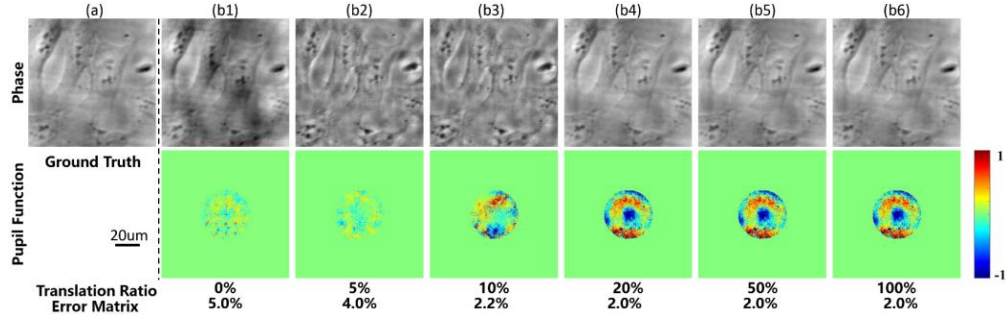

Fig. 9. Pupil recovery improvement strategy with different proportion of translation along 4 axes.

Besides the meniscus-induced wavevector misalignment, the additional aberration makes the recovery of pupil function unstable as shown in Fig. 9(b1), since only 25 BF images are used for recovery, the redundant information is not enough. The black region in Fig. 8(c) is used for an example. And in the meantime, there is also raster grid artifacts [9, 24] (Fig. 9(b1)) due to periodic distribution of LEDs. Instead of creating a custom-design LED array [9, 24], a pupil recovery improvement strategy is presented by utilizing the information of adjacent region to deal with this problem. We assume the adjacent region shares the same pupil function within the coherent area. After 30 iteration of our AWC-FPM algorithm, each tile moves up, down, left, right and return in situ with every 30 iteration. The recovered pupil function is sequentially used as an initial guess for next movement. The number of iterations (i.e. 30 iterations in our experiments) has been tested and does not affect reconstructions. Note that the number of iterations should be enough to make the algorithm converge for every movement. But the proportion of translation has a least value and around 20% movement is required at least as shown in Fig. 9, since the more translation is moved, the more redundant spatial information is generated. Generally, error matrix is used to monitor the performance of recovery [22]. The final error matrix is also listed below each column in Figs. 9(b1)-9(b6). Note that the coherent area is decided by $d \approx 2\lambda h/w = 140$ μm for an LED size $w$=250 μm. The size of each tile should be selected appropriately to ensure the tile and the movement are not larger than the coherent area. Video 1 is provided to illustrate the process. Figure 10 shows the pupil function reconstructions of entire FOV before and after our pupil recovery improvement strategy. The entire FOV of each well will be divided into 7 rows by 10 columns (110 μm×110 μm per tile).

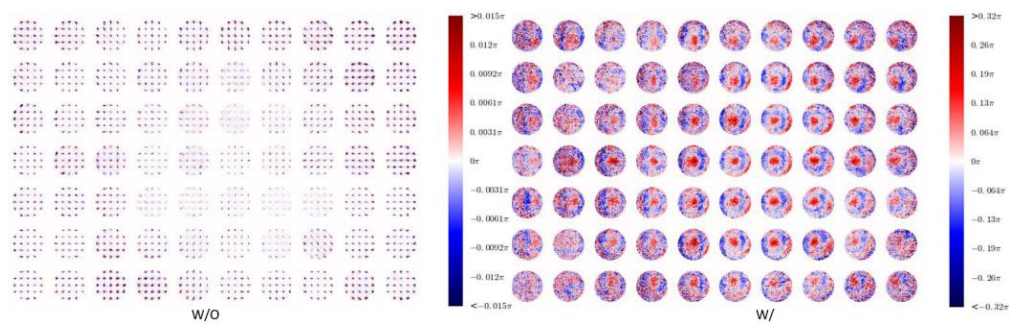

Fig. 10. Pupil function reconstructions of entire FOV. Left: without our pupil recovery improvement strategy; Right: with our strategy.

## 5. Final performances with dual-channel fluorescence imaging

The final performance of the whole system with dual-channel fluorescence imaging is shown in Fig. 11. The montage of all 96 composite images are shown in Fig. 11(a), while a magnified view of random well F12 on the 96-well plate is shown in Fig. 11(b). The close-ups are local aberration recovered by our pupil improvement strategy, LR raw image, phase reconstruction, composite image of dual-channel fluorescence imaging, respectively. The fluorescence images have been deconvolved [25].

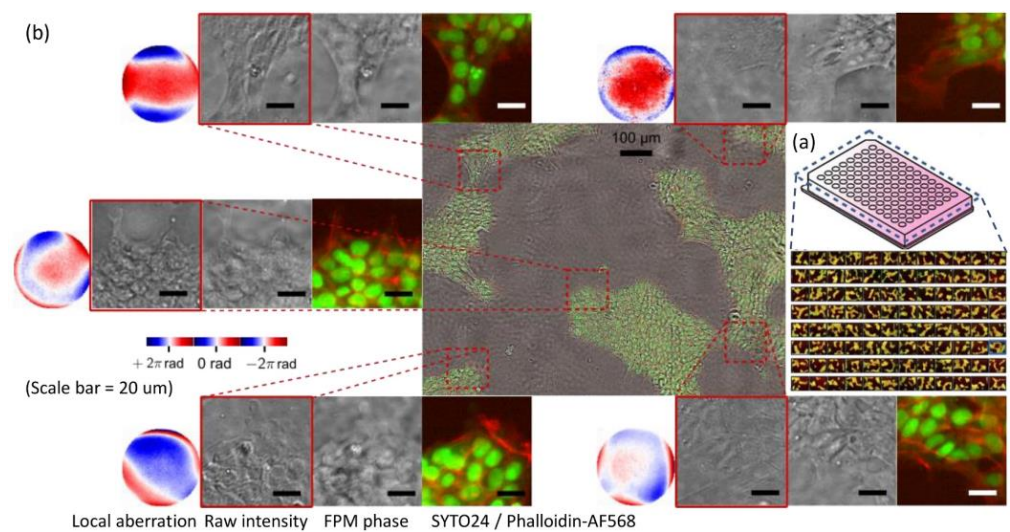

Fig. 11. Final performance of the 96 Eyes with dual-channel fluorescence imaging. (a) Montage of all 96 composite images captured by the 96 Eyes imager. (b) Magnified view of well F12 on the 96-well plate. Close-ups: Recovered aberration function with our improvement strategy, LR raw image, recovered sample phase, fluorescence response from SYTO24 green fluorescent nucleic acid stain and fluorescence response from phalloidin probe conjugated to Alexa Fluor 568 dye, respectively.

The imaging throughput is defined by pixel readout rate (pixel / cell) multiply cell screening rate (cell / second). The statistics of existing commercial products are shown in Fig. 12. The flow cytometer and confocal microscope are two typical extreme examples. A line can be connected from the flow cytometer (legend 2 in Fig. 12) to confocal microscope (legend 11 in Fig. 12) and the higher the throughput is, the farther to top right the location is. To the best of our knowledge, our 96 Eyes system has the largest throughput currently.

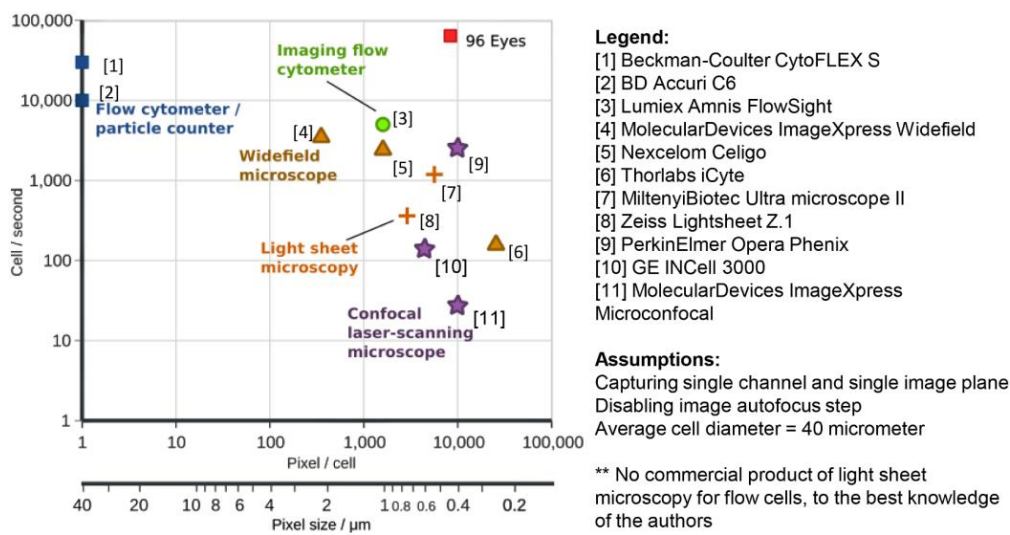

Fig. 12. Comparisons of throughput of existing commercial products.

## 6. Conclusions

In this paper, an AWC-FPM algorithm and a pupil recovery improvement strategy are presented to tackle the meniscus-induced wavevector misalignment and additional aberration. By analyzing the BF images of raw data, a global offset is added to the wavevector of LEDs and a minor shift is compensated with SA modules individually, which solves the misalignment and improves the robustness of FPM algorithm significantly. The effectiveness, validity and performance are verified. Considering only utilizing a few images (25 images) to obtain twice better resolution in our 96 Eyes, the redundant information is not enough to get a robust pupil function. Adjacent spatial information is used to improve the robustness of the recovery of pupil function and the general raster grid disappears. Dual-channel fluorescence excitation is designed to obtain more information for microbiologists. Experiments are demonstrated to verify the final performances of 96 Eyes system. On the one hand, our methods effectively solve the troublesome challenges, the meniscus problems of 96 Eyes in situ and make our system more practical. On the other hand, our algorithms make the FPM algorithm more robust and can be extended to other FPM-based applications to overcome different challenges. We also note the potential of a better imaging robustness through hardware improvement. For instance, meniscus-free culture plates [26] can be adopted to our 96 Eyes to tackle the model misalignment, thus reduces computational burden in turn. We anticipate the 96 Eyes will accelerate and help biomedical and pharmaceutical research that utilizes high-throughput cell imaging and screening format, for example, the cervical cancer screening.

## Disclosures

The authors have no relevant financial interests in this article and no potential conflicts of interest to disclose.

## Funding

Caltech Agency Award (AMGEN.96EYES).

## Acknowledgement

We would like to thank Albert Chung and Daniel Martin (California Institute of Technology, USA) for helpful discussions.